\documentclass[pre,twocolumn,floatfix]{revtex4}

\usepackage{graphicx}
\usepackage{amsmath}

\begin{document}

\newcommand{\nF}{\, nF}
\newcommand{\mV}{\, mV}
\newcommand{\mS}{\, mS}
\newcommand{\muS}{\, \mu S}
\newcommand{\ms}{\, ms}
\newcommand{\s}{\, s}
\newcommand{\E}{\mathbf E}

\title{Spatial representation of temporal information through spike timing
  dependent plasticity}

\author{Thomas Nowotny}
\email{tnowotny@ucsd.edu}
\homepage{http://inls.ucsd.edu/~nowotny}
\affiliation{Institute for Nonlinear Science, University of California San
  Diego, 9500 Gilman Drive, La Jolla, CA 92093-0402}
\author{Misha I. Rabinovich}
\email{mrabinovich@ucsd.edu}
\affiliation{Institute for Nonlinear Science, University of California San
  Diego, 9500 Gilman Drive, La Jolla, CA 92093-0402}
\author{Henry D.~I.~Abarbanel}
\email{hdia@jacobi.ucsd.edu}
\affiliation{Institute for Nonlinear Science, Department of Physics \\ and \\
  Marine Physical Laboratory (Scripps Institution of Oceanography),
  University of California, San Diego, La Jolla, CA 92093-0402}

\begin{abstract}
  We suggest a mechanism based on spike time dependent plasticity (STDP) of
  synapses to store, retrieve and predict temporal sequences. The mechanism
  is demonstrated in a model system of simplified integrate-and-fire type
  neurons densely connected by STDP synapses. All synapses are modified
  according to the so-called normal STDP rule observed in various real
  biological synapses. After conditioning through repeated input of a limited
  number of of temporal sequences the system is able to complete the temporal
  sequence upon receiving the input of a fraction of them. This is an example
  of effective unsupervised learning in an biologically realistic system. We
  investigate the dependence of learning success on entrainment time, system
  size and presence of noise. Possible applications include learning of motor
  sequences, recognition and prediction of temporal sensory information in
  the visual as well as the auditory system and late processing in the
  olfactory system of insects.
\end{abstract}

\maketitle

\section{Introduction}\label{intro}

Animals are challenged in various ways to learn, produce, reproduce and
predict temporal patterns. A prominent example are the numerous motor programs
necessary to interact efficiently with the environment. One specific
manifestation is the vocal motor system of song birds. It has been shown that
the temporal sequence of syllables in a bird's song corresponds to temporal
sequences of bursts in the neurons of the forebrain control
system~\cite{Hahnloser2002,Dave1998,Yu1996}. These are learned and stored by the adolescent
bird.

Temporal codes seem to be used for a variety of other tasks as well.
Temporal coding in the retina~\cite{Rullen2001} is an example, as is
information transport in the olfactory system of the locust. In the latter it
has been shown that the purely identity coded information of the receptor
neurons is transformed into an identity-temporal code inside the antennal
lobe~\cite{Laurent1998,Laurent1999,Laurent2001}.

Whereas there is a long history of research on sequence learning and recognition in the
framework of abstract neural networks (cf the relevant chapters in
\cite{Hertz1991} and \cite{Trappenberg2002} and references therein) it is an
open question how the learning and memory of time sequences is accomplished
in real biological neural systems. Three main principles for representing
time in neural systems are frequently discussed:
\begin{itemize}
\item The first makes use of delays and filters. There are various ways of
processing of temporal information in the dendritic
tree~\cite{Vetter2001,Haeusser2000,Koch1999,Koch1984} or through axonal
delays~\cite{Jeffres1948,Sompolinsky1986,Herz1988,Herz1989,Gerstner1993,Hertz1996,Leibold2002}. Other
examples are multilayer neural networks in which the delay of the synaptic
connections between layers allows to represent or decode temporal information
and propagating waves as known from the thalamo-cortical
system~\cite{Rinzel1998,Colomb1996}.

\item The second principle rests on feedback. Through delayed feedback temporal
information can be processed on the level of individual neurons as well as on
the level of larger structures. A prominent example for this are recurrent
multi layer neural networks which play a role in sequence memory in the
hippocampus~\cite{Lisman2001,Lisman1999}.

\item The third principle is to transform the temporal information into
spatial information. This can occur through the dynamics of a network with
asymmetric lateral inhibition~\cite{Tonkin1996}.
\end{itemize}
In this paper we demonstrate an alternative mechanism which maps the temporal
information to the strength of synapses in a network through spike timing
dependent plasticity (STDP). Similar mechanisms have been suggested for
predictive activity and direction selectivity in the visual
system~\cite{Rao2000} and learning in the
hippocampus~\cite{Abbott1996,Lisman2001,Lisman1999} as well as prediction in
hippocampal place fields and route learning in
rats~\cite{Mehta1997,Gerstner1997,Jensen1996}. In contrast to these earlier
works we focus on questions
of learning of several distinct input sequences in one system and a sparse
coding scheme. This learning
capability is necessary in order to process the identity-temporal code
believed to be generated by winnerless competition in sensory
systems~\cite{Laurent2001,Rabinovich2001}.

Synaptic plasticity in the connections among neurons allows networks to alter
the details of their interaction and develop memories of previous input
signals. The details of the methods by which biological neurons express
plasticity at synapses is not fully understood at the biophysical level, but
many aspects of the phenomena which occur when presynaptic and postsynaptic
neurons are jointly activated are now becoming clear. First of all, it seems
well established that activity at both the presynaptic and the postsynaptic
parts of a neural junction is required for the synaptic strength to
change. Arrival of a presynaptic action potential will induce, through normal
neurotransmitter release and reception by postsynaptic receptors, a
postsynaptic electrical action which generally leads to no change in the
coupling strength at that synapse. Depolarization of the postsynaptic cell by
various means {\em coupled with} arrival of a presynaptic action potential
can lead to changes in synaptic strength in a variety of experimental
protocols.  It is quite important that changes in the synaptic strength,
which we denote in terms of a conductivity change $\Delta g$ can be either
positive, called potentiation, or negative, called depression. When the
expression of $\Delta g$ is long lasting, several hours or even much longer
after induction, increases in $g$ are called long term potentiation or LTP,
and decreases in $g$ are called long term depression or LTD. Good reviews of
the current situation are found in~\cite{malenka1999,linden1999,Bi2001}.

LTP and LTD can be induced by (1) depolarizing the postsynaptic cell to a
fixed membrane voltage and presenting presynaptic spiking activity at various
frequencies, by (2) inducing slow (LTD) or rapid (LTP) release of
$\text{Ca}^{2+}$~\cite{Yang1999}, or by (3) activating the presynaptic terminal
a few tens of milliseconds before activating the postsynaptic cell, leading to
LTP, or presenting the activation in the other order, leading to
LTD~\cite{Markram1997,Bi1998}.

In this paper we study numerically a network composed of integrate-and-fire
neurons which are densely
coupled with synaptic interactions whose maximal conductances are permitted
to change in accordance with the observations on closely spaced spike arrival
times to the presynaptic and postsynaptic junctions of the synapse.

\begin{figure}
  \includegraphics[width=0.8\columnwidth]{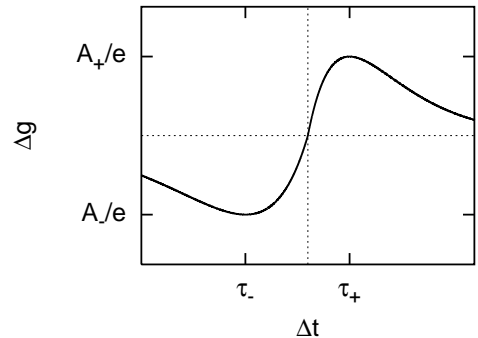}
  \caption{\label{figure1} STDP learning rule. $\Delta g= A_+
      \frac{\Delta t}{\tau_+} e^{- \Delta t / \tau_+}$ for $\Delta t > 0$ and
      $\Delta g= A_- \frac{\Delta t}{\tau_-} e^{ \Delta t / \tau_-}$ for
      $\Delta t < 0$, $A_+$, $A_- > 0$. This form of the learning rule was
      directly inferred from experimental data \cite{Bi1998}}
\end{figure}

The response of a learning synapse to the arrival of a presynaptic spike at
$t_{\text{pre}}$ and a postsynaptic spike at $t_{\text{post}}$ is a function
only of $\Delta t = t_{\text{post}} - t_{\text{pre}}$ and for $\Delta t > 0$
$\Delta g(\Delta t)$ is positive, LTP, and for $\Delta t < 0$
$\Delta g(\Delta t)$ is negative, LTD. 
\begin{align} \Delta g(\Delta t) &= A_+
\frac{\Delta t}{\tau_+} e^{-\frac{\Delta t}{\tau_+}} \quad \text{for} \quad
\Delta t > 0 \nonumber \\ \Delta g(\Delta t) &= A_- \frac{\Delta t}{\tau_-}
e^{\frac{\Delta t}{\tau_-}} \quad \text{for} \quad \Delta t < 0
\label{learneqn}
\end{align}
where $A_+$, $A_-$, $\tau_+$ and $\tau_-$ are positive constants (see Fig.\
\ref{figure1}). Synaptic plasticity of this type is often referred to as
spike timing dependent plasticity (STDP).  For many mammalian {\em in vitro}
or cultured preparations the characteristic LTD time $\tau_-$ is about two or
three times longer than the characteristic LTP time $\tau_+$.

Here we inquire how a network composed of familiar integrate-and-fire neurons
can develop preferred spatial patterns of connectivity when interacting
through synapses which update their strength according to the STDP learning
rule just given. This rule is a simplification, which applies for our setting
of spiking neurons, of more general
models~\cite{Lisman1989,Holmes1990,Zador1990,Gold1994,Schiegg1995,Shouval2002,Abarbanel2002}
which indicate how $\Delta g(\Delta t)$ behaves under stimulus of arbitrary
presynaptic and postsynaptic waveforms.

The transformation of temporal information into synapse strength through STDP
maps a temporal sequence of excitations of neurons to a chain of stronger or
weaker synapses among these neurons. If the synapses are excitatory, a
strengthened chain of synapses facilitates subsequent excitations of the same
temporal pattern up to a point that activation of a few neurons from the
temporal sequence allows the system to complete the remaining sequence. The
temporal sequence thus has been learned by the system.  We demonstrate this
type of sequence learning in a computer simulation of a system with
integrate-and-fire neurons and Rall type synapses and investigate the
reliability of learning, the storage capacity in terms of the number of
stored sequences, the scaling of both with system size and sequence length
and the robustness against different types of noise.

\section{Model system} 

\subsection{Components and Connections} 

\begin{figure} 
\includegraphics[width=\columnwidth]{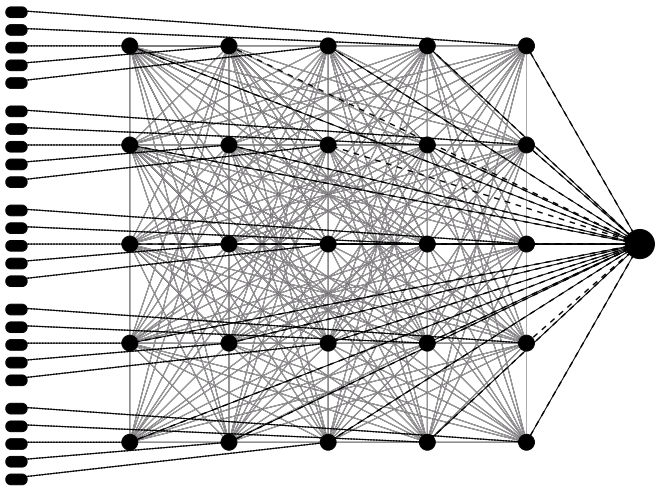}
\caption{\label{figure2} Morphology of the model system. The ovals are
    artificial input neurons producing rectangular spikes of $3 \ms$ duration
    at specified times. Each is connected by a non-plastic excitatory synapse
    to one of the neurons in the main ``cortex'' (dotted lines). The full
    circles depict the integrate-and-fire neurons of the main
    ``cortex''. They are connected all-to-all by STDP synapses shown as
    solid gray lines. The big full circle on the right depicts a neuron with
    slow Calcium dynamics which inhibits all neurons in the ``cortex''
    through the non-plastic synapses shown as dashed lines.}
\end{figure}

\begin{figure}
  \includegraphics[width=\columnwidth]{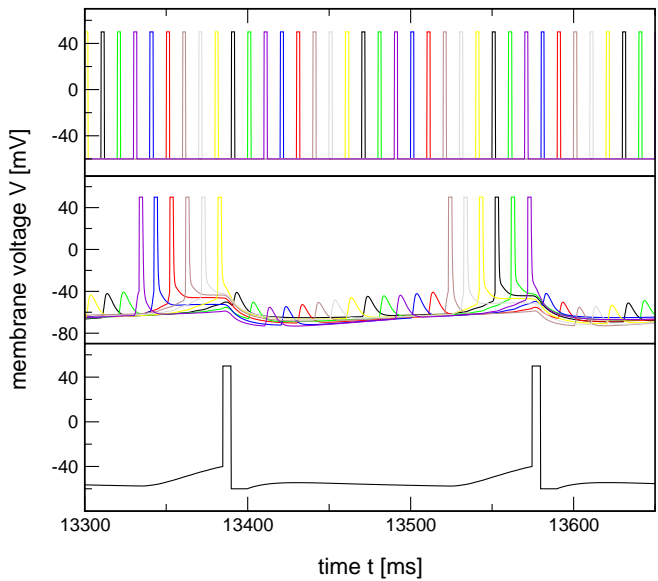}
  \caption{\label{figure3} Typical piece of a training session. The
  rectangular spikes in the upper panel are the input signal spaced by
  $10\ms$ in this example. The traces in the middle panel are the
  integrate-and-fire memory neurons. The slow spike train in the bottom panel
  belongs to the globally inhibitory neuron. Note the instantaneous onset of
  the spikes in the integrate-and-fire neurons and how the inhibitory neuron
  segments the input into pieces of 6 spikes each.}
\end{figure}

To explore the learning principle we simulated a
network with the topology shown in Fig.\ \ref{figure2}.  In this network $n$
integrate-and-fire neurons are connected all-to-all while each neuron
also receives input from one ``input neuron'' (filled ovals in Fig.\
\ref{figure2}).

The ``input neurons'' generate rectangular spikes of $3 \ms$ duration at
times determined by externally chosen input sequences. Each of these spikes
is sufficient to trigger exactly one spike in the receiving neuron (see Fig.\
\ref{figure3}). The input sequences are chosen such that only one ``input
neuron'' spikes at any given time and the time between input spikes was fixed
in the normal test setup. In section \ref{noisesec} these input neurons are
replaced by Poisson neurons with random spike times.

The membrane voltage of the integrate-and-fire neurons used in this study
for sub-threshold activity is given by
\begin{align}
  C \frac{d V}{dt} = -g_{\text{Leak}}(V(t)-V_{\text{Leak}})
  +I_{\text{Synapse}}(t) \label{ifeqn}
\end{align}
where $C= 0.2 \nF$, $g_{\text{Leak}} = 0.3 \muS$ and $V_{\text{Leak}}= -60
\mV$. Whenever the membrane potential $V(t)$ reaches $V_{\text{th}} = -40
\mV$, it is set to the firing voltage $V_{\text{max}}= 50 \mV$, kept at that
voltage for $t_{\text{fire}}= 2 ms$ and then released into the normal
integration state. The neuron is subsequently refractory for
$t_{\text{refract}} = 40 ms$ before another firing event is allowed. During
the refractory period the neurons integrates normally but the transition of
the firing threshold has no effect. In the implementation of
integrate-and-fire neurons used in this work no crossing of the firing
threshold from below is necessary to elicit a spike in a super-threshold
neuron after the refractory period. See Fig.\ \ref{figure3}, middle panel,
for typical spike forms.
    
A neuron connected to all neurons in the network (large filled circle in
Fig.\ \ref{figure2}) provides global inhibition whenever the activity in the
network exceeds a certain threshold. The inhibitory neuron is an
integrate-and-fire neuron governed by (\ref{ifeqn}) with $C= 1.0 \nF$,
$g_{\text{Leak}}= 0.01 \muS$, $V_{\text{Leak}}= -60 \mV$, $V_{\text{th}}= -40
\mV$, $V_{\text{max}}= 50$ and $t_{\text{fire}}= 5 \ms$. In contrast to the
memory neurons this neuron is reset to its resting potential
$V_{\text{Leak}}$ after each firing. Then the membrane potential is fixed to
$V_{\text{Leak}}$ for $t_{\text{refract}}= 10 \ms$ until normal integration
resumes. The inhibitory neuron was implemented as a resetting integrate-and
-fire neuron because it has a very weak leak current allowing integration
over long time windows. This weak leak current would cause very unnatural
broad spikes in a non-resetting neuron. A typical voltage trace is shown in
the lowest panel of Fig.\ \ref{figure3}.
  
Our model of the synapses comes from Rall~\cite{Rall1967, Rall1989} and now
is a standard model for simplified synaptic dynamics~\cite{Destexhe1998}. In
particular, we use
\begin{align}
I_{\text{Synapse}} = - g_{\text{syn}} \, g(t) \,
(V_{\text{post}}(t) - V_{\text{syn}}) ,
\end{align}
where $g(t)$ satisfies
\begin{align}
  \frac{d f(t)}{d t} &= \frac{1}{\tau_{\text{syn}}} \,
  (\Theta(V_{\text{pre}}(t) - V_{\text{th}}) - f(t)) \nonumber \\
  \frac{d g(t)}{d t} &= \frac{1}{\tau_{\text{syn}}} \, (f(t) - g(t)) ,
\end{align}
and $V_{\text{syn}}= 0 \mV$, $V_{\text{th}}= -20 \mV$, $\tau_{\text{syn}}= 15
\ms$, $V_{\text{pre}}(t)$ and $V_{\text{post}}(t)$ are the pre- and
postsynaptic membrane potentials and $g_{\text{syn}}$ is the strength of the
synapse. $\Theta(u) = 0, u \le 0$ and $\Theta(u) = 1, u>0$ is the usual
Heaviside function. Typical EPSPs generated by these synapses can be seen in
the middle panel of Fig.\ \ref{figure3}.

The synaptic strength of the internal synapses is adjusted according to the
synaptic plasticity rule shown in in Fig.\ \ref{figure1} whenever a spike in
their presynaptic and postsynaptic neuron occurs. In itself, this rule may
lead to ``run-away'' behavior of the synaptic strengths. While this may be
avoided in the dynamical model of synaptic plasticity~\cite{Abarbanel2002},
we need to address this within the simpler model used here. We do so by
two approaches: (1) we add a long term, slow decay to the synaptic plasticity
which would, all other factors being absent, bring it back to a nominal
allowed level a long time after alteration by our rule. This we implement
with
\begin{align}
  \frac{d{g}_{\text{raw}}}{dt}= - \frac{1}{\tau_g}
  (g_{\text{raw}}(t)-g_{0,\text{raw}})
\end{align}
where $g_{0,\text{raw}}$ is the initial value of the unmodified synapse
strength. So after potentiation or depression according to the synaptic
plasticity rule, the synaptic strength is allowed to slowly decay back to its
original value. The time scale of this exponential decay is set by $\tau_g=
200 \s$. (2) ${g}_{\text{raw}}$ is an intermediate variable which is then
translated into the synaptic strength $g_{\text{syn}}$ via a sigmoid
saturation rule
\begin{align}
  g_{\text{syn}}= g_{\text{max}} {\textstyle \frac{1}{2}} \big(\tanh
  \big(g_{\text{slope}}(g_{\text{raw}} - g_{1/2})\big) + 1\big),
\end{align}
where $g_{\text{max}}$ is the largest allowed value for the synaptic
conductivity, and $g_{1/2}$ sets the threshold where saturation to this value
is implemented. All data shown in this work was obtained with
$g_{\text{max}}= 2.8 \muS$, $g_{1/2} = 1/2 \, g_{\text{max}}$ and
$g_{\text{slope}} = 1/g_{1/2}$. In addition the globally inhibitory neuron
tends to curb the tendency of the network to saturate its synaptic strengths.

These features of our model reflect our lack of knowledge of the biophysical
factors setting the synaptic strength in the first place and our equivalent
lack of knowledge how these factors bound the eventual rise or fall of
synaptic strength. Our assumption in using these rules is that the actual
mechanisms, while surely more complicated in detail, will provide the same
effective bounding feature.

The complete system is realized in C++ using an order $5$/$6$ variable time
step Runge-Kutta algorithm~\cite{Butcher1987}. The error goal per time step
was $10^{-7}$ in all simulations. A run of $100$ simulated seconds of a
system with 50 neurons takes about $3$ hours on an Athlon $1.4$ GHz
processor.

This model system mimics the situation of a highly connected piece
of cortex receiving input from the neural periphery. Our input
can be interpreted in two ways. It might be a single strong excitatory
postsynaptic potential (EPSP) received from an upstream neuron which is
strong enough to trigger a spike. It could also be interpreted as the
coincidence of several weaker EPSPs received from various presynaptic
neurons being sufficient to cause a spike.

\subsection{Operations and Activity}

To test the ability of this network to store (learn) and retrieve (remember)
temporal-identity patterns it was trained with sets of randomly chosen
sequences of inputs. These sequences were chosen without repetition of
neurons within the sequence. Note that this implies a minimal time of the
order of the length of the sequence between spikes in each neuron. For this reason the
choice of resetting or non-resetting neurons is not important as the
integration times of the neurons are small compared to the total length of
the sequences and the time scale of the global inhibition. Our choice of
non-resetting integrate-and-fire neurons was mainly guided by the more
natural spike form of the non-resetting variety.

The sequences were presented continuously with the first neuron of the
sequence following the last with the same time delay as the neurons within
the sequence. The global inhibition of the system partitions this continuous
input of spikes into pieces of about $6-8$ spikes at a time. Between these
input windows the
whole system is inhibited and thus reset. This mechanism can be seen in the
example training session shown in Fig. \ref{figure3}. Note that the details
of the global inhibition mechanism do not matter as long as the system is
efficiently reset after an appropriate amount of activity.

The learning rate $A_+$ and the time scale of forgetting
$\tau_g$ in the synaptic plasticity learning rule were chosen such that
learning reaches a steady state after a learning time of about $1600 \Delta
t$, where $\Delta t$ is the fixed inter spike interval between input
activations. For an example of the learning protocol see Fig.\
\ref{figure3}. In all studies described below $\Delta t$ was chosen as
$\Delta t= 10 \ms$, $15 \ms$ or $20 \ms$. The learning rule has to accommodate
all these input speeds and possibly values in between. In particular we here
chose $A_+= 0.3 \muS$, $A_-= 2/3 \, A_+$, $\tau_+ = 16 \ms$, $\tau_- = 3/2 \,
\tau_+$ and $\tau_g = 200 \s$.

After the training phase the network was presented with pieces of the
training patterns. We presented all possible ordered pieces of one to four
input spikes and recorded the number and identity of spiking neurons in the
network in response to this input. Perfect learning of the patterns would
correspond to obtaining a spike from each of the network neurons in a given
pattern when presenting a piece of two or three inputs from that pattern to
the ``input neurons''.  Furthermore, all other network neurons should remain
inactive if the pattern is reproduced exactly.

As a result of incomplete or ineffective learning two types of errors can
occur. (1) Neurons which should be excited within the given pattern do not
spike or (2) neurons which are not supposed to spike do so. Due to overlap of
input patterns, the learning efficiency is a function of the number of
learned patterns as well as the size of the network. Therefore, estimating
the expected amount of overlaps in the randomly chosen input sequences
provides information about the optimally achievable system performance.

The probability distribution for the number $Y_{ijrkn}$ of {\em ordered}
$j$-tuples occurring in at least $i$ out of $r$ patterns with $k$ neurons
each for a system with a total number of $n$ neurons can be calculated in the
following way: First consider a given ordered $j$-tuple and a given pattern
with $k$ neurons. The sequence is presented continuously and therefore needs
to be interpreted as cyclically closed. Thus there are $k$ possibilities to
position the $j$ tuple in the sequence (starting at neuron $1$ to starting at
neuron $k$) and $(n-j)! / (n-j-(k-j))!$ possibilities to choose the remaining
neurons in the sequence. The total number of sequences of length $k$ is $n! /
(n-k)!$. Therefore, the probability $p_j$ to have a given {\em ordered}
$j$-tuple in a given pattern with $k$ active neurons is given by
\begin{align}
  p_j= k \frac{(n-j)!}{(n-k)!} \Big/ \frac{n!}{(n-k)!} = k \frac{(n-j)!}{n!}.
\end{align}
If $r$ sequences of length $k$ are chosen independently, the probability
to have any given {\em ordered} $j$-tuple of neurons in $i$ or more of the
$r$ sequences is given by the binomial distribution with parameters $r$ and
$p_j$,
\begin{align}
  p_j^i= \sum_{s= i}^{r} \binom{r}{s} (p_j)^s (1-p_j)^{r-s}.
\end{align}
In good approximation one can treat the events of one given $j$-tuple being
in $i$ or more sequences and another $j$-tuple being so as independent. In
this approximation the probability distribution for $Y_{ijrkn}$ is again a
binomial distribution with parameters $\frac{n!}{(n-j)!}$ and $p_j^i$, 
\begin{align}
  P(Y_{ijrkn} = l) \approx \binom{\frac{n!}{(n-j)!}}{l} (p_j^i)^l
  (1-p_j^i)^{\frac{n!}{(n-j)!}-l} . \label{Ydistri}
\end{align}

\begin{figure}
  \includegraphics[width=\columnwidth]{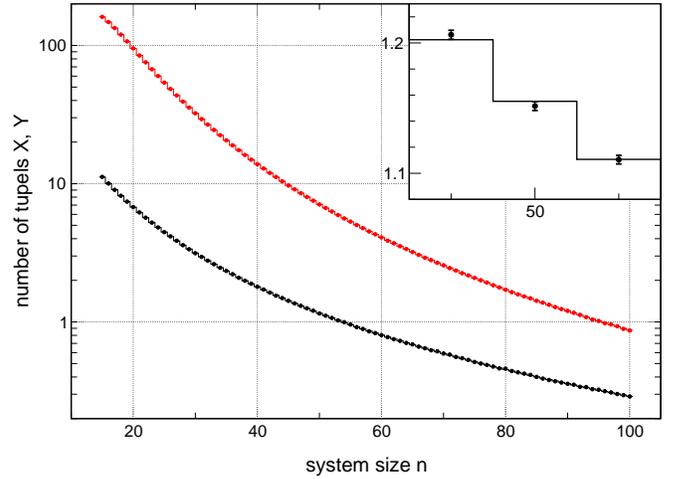}
  \caption{\label{figure4} Comparison of the expectation values for
    $Y_{2,2,10,8,n}$ (lower line) and $X_{3,2,10,8,n}$ (upper line) obtained
    from (\ref{Ydistri}) and (\ref{Xdistri}) to the normalized number of
    occurrences of unordered $3$-tuples (gray dots) and ordered $2$-tuples
    (black dots) in more than $2$ sequences in $100000$ randomly generated
    sets of $10$ sequences of length $8$. The inlay shows a closeup of the
    data on ordered tuples in the region with system size around $50$ neurons
    which is the size used in most numerical simulations.}
\end{figure}

Fig.\ \ref{figure4} shows a comparison of the expectation value for $\E
Y_{2,2,10,8,n}$ obtained from this approximate distribution compared to the
relative number of occurrences in $100000$ randomly generated sets of $10$
sequences of length $8$. There is no significant discrepancy which
demonstrates the precision of the estimate.

The probability distribution for the number $X_{ijrkn}$ of {\em unordered}
$j$-tuples occurring in at least $i$ out of $r$ patterns with $k$ neurons
each for a system with a total number of $n$ neurons can be calculated pretty
much in the same way. Now the the probability $\hat{p}_j$ to have a given
{\em unordered} $j$-tuple in a given pattern with $k$ active neurons is
\begin{align}
  \hat{p}_j= \binom{n-j}{k-j} \Big/ \binom{n}{k} .
\end{align}
Then, the probability $\hat{p}_j^i$ to have any given {\em unordered} $j$-tuple
of neurons in $i$ or more of $r$ independently chosen patterns is the
binomial distribution with parameters $r$ and $\hat{p}_j$,
\begin{align}
\hat{p}_j^i= \sum_{s= i}^{r} \binom{r}{s} (\hat{p}_j)^s (1-\hat{p}_j)^{r-s} .
\end{align}
Again taking the approximation of assuming independence for the occurrence of
distinct tuples this leads once more to a binomial distribution, now
with parameters $\binom{n}{j}$ and $\hat{p}_j^i$,
\begin{align}
  P(X_{ijrkn} = l) \approx \binom{\binom{n}{j}}{l} (\hat{p}_j^i)^l
  (1-\hat{p}_j^i)^{\binom{n}{j}-l}. \label{Xdistri}
\end{align}
The comparison of the expectation values $\E X_{2,3,10,8,n}$ with
numerically observed relative numbers of occurrence in Fig.\ \ref{figure4}
shows again a perfect match.

The model parameters were chosen such that two to three spiking predecessors
of a given neuron in a trained sequence are sufficient to excite that
neuron. The learning performance is therefore poor as long as there is a
significant amount of ordered $2$-tuple overlaps in the patterns. The rule of
thumb $\E Y_{22rkn} < 0.5$ for the expectation value of $Y_{ijrkn}$, provides
an estimate for number $r$ of pattern of length $k$ that can be successfully
stored in a system of $n$ neurons. Another estimate for the number
of learnable sequences is provided by the rule of thumb $\E X_{23rkn} < 0.5$,
i.e.\ the overlaps in input sequences should have negligible impact on the
learning if there is no significant amount of {\em unordered} $3$-tuples
occurring in more than one pattern.

Typically capacity estimates are given in the limit of the system size $n$
tending to infinity. As shown in Appendix \ref{appe} the leading term of the
Taylor expansion of $p_j^i$ with respect to $p_j$ around $p_j= 0$ is
\begin{align}
  p_j^i = \binom{r}{i} (p_j)^i + {\cal O}((p_j)^{i+1})
\end{align}
such that the asymptotic equation
\begin{align}
  \lim_{n \to \infty} \E Y_{ijrkn} \stackrel{!}{=} \epsilon
\end{align}
leads to
\begin{align}
  & \lim_{n \to \infty} \, \frac{n!}{(n-j)!} \binom{r}{i}
  \Big(k\frac{(n-j)!}{n!}\Big)^i = \epsilon \\ 
  \Longleftrightarrow \quad & \lim_{n \to \infty} \, \frac{r^i}{i!} k^i
  n^{-j(i-1)} = \epsilon
\end{align}
such that the capacity $r(n,k,\epsilon)$ is asymptotically
\begin{align}
  r(n,k,\epsilon) = \frac{1}{k} (i! \epsilon)^{\frac{1}{i}}
  n^{\frac{j(i-1)}{i}} .
\end{align}
In the same way
\begin{align}
    \lim_{n \to \infty} \E X_{ijrkn} \stackrel{!}{=} \epsilon
\end{align}
leads to
\begin{align}
  \hat{r}(n,k,\epsilon) = \frac{(k-j)!}{k!} (i! j! \epsilon)^{\frac{1}{i}}
  n^{\frac{j(i-1)}{i}} .
\end{align}

\begin{figure}
  \includegraphics[width=\columnwidth]{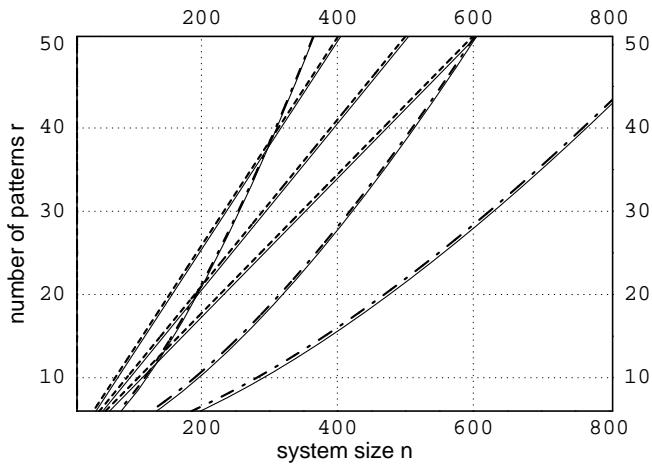}
  \caption{\label{figure5} Estimate for the maximum storage capacity of the
    system. The dashed lines divide the plane into two regions with $\E
    Y_{2,2,r,k,50} > 0.5$ above and $\E Y_{2,2,r,k,50} < 0.5$ below for $k=
    8$ (topmost line), $10$ (middle line), and $12$ (lowest line)
    respectively. The thin solid lines are the corresponding estimates for
    the asymptotically correct values $r(50,k,\frac{1}{2})$. The dash-dotted
    lines analogously mark the boundaries between regions with $\E
    X_{2,3,r,k,50} > 0.5$ above and $\E X_{2,3,r,k,50} < 0.5$ below. Again
    the thin lines are the asymptotically correct estimates
    $\hat{r}(50,k,\frac{1}{2})$}
\end{figure}

The dashed lines in Fig.\ \ref{figure5} are some examples for the first rule
of thumb $\E Y_{22rkn} = \frac{1}{2}$ and the thin solid lines are the
corresponding values of $r(n,k,\frac{1}{2})$.  The estimates based on the
rule $\E X_{23rkn} = \frac{1}{2}$ are shown as dash-dotted lines in Fig.\
ref{figure5} and the corresponding values of the asymptotically correct
$\hat{r}(n,k,\frac{1}{2})$ are again shown as thin solid lines. The
correspondence between the exact evaluation of the capacity estimators and
the asymptotically correct capacity functions $r(n,k,\epsilon)$ and
$\hat{r}(n,k,\epsilon)$ is noteworthy. The relative capacities $r'(k) := k
r(n,k,\epsilon)/ n^{\frac{j(i-1)}{i}} = (i! \epsilon)^{\frac{1}{i}}$ and
$\hat{r}'(k) := k \hat{r}(n,k,\epsilon)/n^{\frac{j(i-1)}{i}} =
\frac{(k-j)!}{(k-1)!} (i!j!\epsilon)^{\frac{1}{i}}$ behave quite
differently. Whereas the former is constant with respect to $k$ the latter is
falling in $k$. So, depending whether a system is more sensitive to ordered
tuple overlaps or to unordered tuple overlaps, the relative capacity is
constant or falling in $k$. In particular for systems sensitive to unordered
tuple overlaps it will be beneficial to store many short sequences instead of
a few long ones.

\section{Results}

\begin{figure}
  \begin{center}
    \includegraphics[width=0.7\columnwidth]{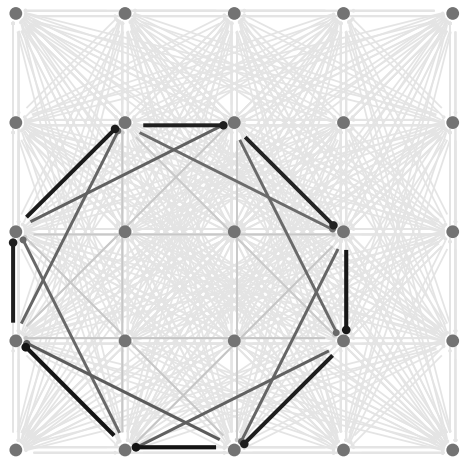}
  \end{center}
  \caption{\label{figure6} Simple example of a learned identity-temporal
    pattern. The neurons at the corners of the octagon have been repeatedly
    excited in clockwise order. The width and grayscale of the connections
    encodes the strength of the corresponding synapse and the small circle at
    the end shows its direction. As one clearly can see the temporal pattern is
    transformed into an ordered spatial pattern by synaptic
    plasticity.}
\end{figure}

The synaptic plasticity of synapses transforms time sequences of excitation
of neurons into directed spatial patterns as intended. A simple example is
shown in Fig.\ \ref{figure6} for one input pattern. For randomly chosen input
sequences the patterns are structured in the same way but are not so easy to
detect with the human eye.

\begin{figure}
\includegraphics[width=\columnwidth]{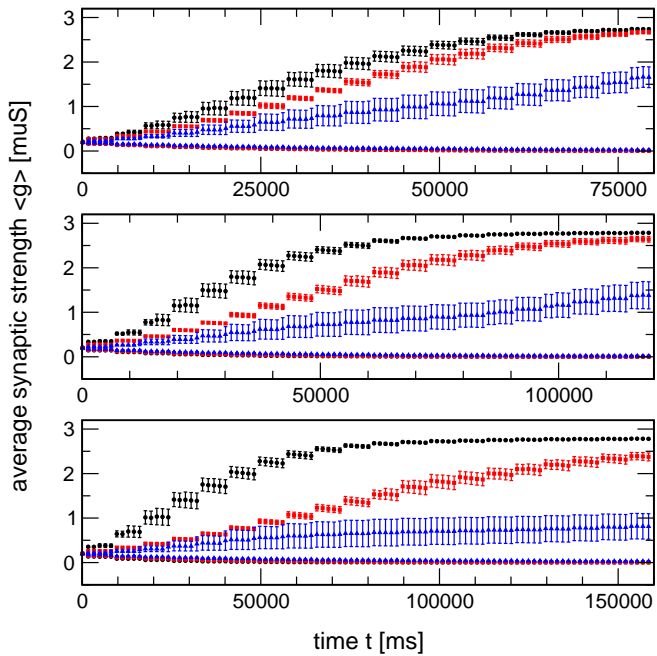}
  \caption{\label{figure7}
    Development of synaptic strength during training. The network of $50$
    neurons was trained with $5$ sequences of length $8$ in sequential order.
    The topmost panel shows the data for sequences entrained with inter spike interval $\Delta
    t= 10 \ms$, the middle with $\Delta t= 15 \ms$, and the lowest with
    $\Delta t= 20 \ms$. Each sequence was presented for $80 \Delta t$ at a
    time. The data shown are average synaptic strengths of synapses between
    the neurons of one of the trained sequences. The topmost points are the
    average strengths of all synapses between the neurons and their direct
    successors in the sequence, the middle are the corresponding strengths of
    synapses between neurons who are next nearest neighbors in the sequence
    under consideration, and the lower points correspond to strengths of
    synapses between neurons with distance $3$ in the sequence.
    The lowest data points are the strengths between the neurons of the
    sequence as described above but {\em against} the order of activation in the
    trained sequence. The sharp rises
    in synaptic strength correspond to training of the particular sequence
    shown here and the falling flanks correspond to the decay while other
    patterns are trained.}
\end{figure}
During training the synapses between consecutively active neurons are
strengthened if pointing in the direction of the activation order or weakened
if connecting the neurons in the wrong direction. An example of the
development of the average synaptic strength of synapses between neurons of
one out of $5$ trained sequences is shown in Fig.\ \ref{figure7}. Note that
the time course and final strength of the synapses depends on the speed with
which the sequences are entrained due to the non-constant learning curve
(\ref{learneqn}).

The ability to store more than one pattern was tested in various setups. We
mainly varied choice, number and length of input sequences and the
speed of entrainment.

\begin{figure}
\includegraphics[width=\columnwidth]{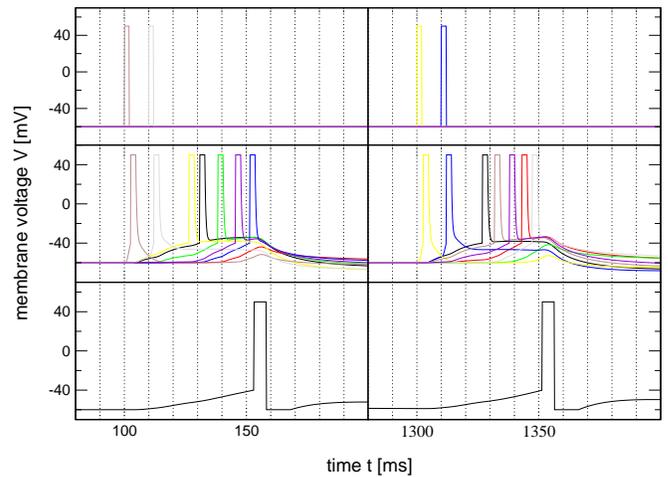}
  \caption{\label{figure8} Typical recall episodes. The system of $50$
    neurons was trained with $2$ (left panel) or $5$ (right panel) sequences
    for $1600 \Delta t$ per sequence, where $\Delta t= 10 \ms$. It then
    receives a cue of two spikes from one of the trained sequences and
    autonomously completes the sequence until stopped by the globally
    inhibitory neuron. Note that although the recall of the identity and
    order of the neurons is perfect in both cases, the exact timing is
    lost. In general one sees a tendency of speed-up to the end of the
    recalled sequence. This can have the effect of destroying the correct
    order of recall in the later sequence if the global inhibition is not
    present.}
\end{figure}

\begin{figure}
\includegraphics[width=\columnwidth]{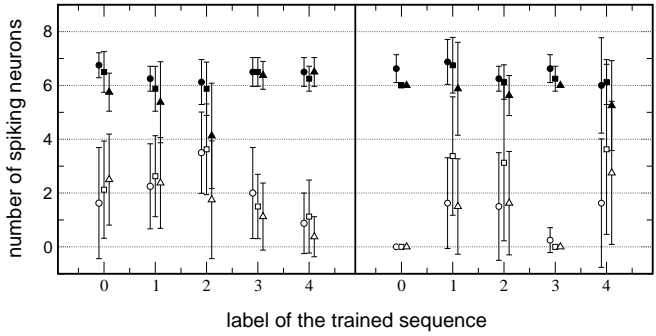}
 \caption{\label{figure9} Examples of learning in a $50$ neuron network after
    $1600 \Delta t$ sequential training with $5$ input sequences of length
    $8$. The left and the right panel show results for two independently
    chosen sets of $5$ input sequences labeled with numbers 0 to 4 in each
    set. The filled symbols show the average number of spiking neurons within
    a tested sequence and the open symbols erroneously spiking neurons. The
    test cue were fractions of length $2$ from the trained sequences. The
    circles were obtained with a training speed of $\Delta t= 10 \ms$, the
    squares with $\Delta t= 15 \ms$, and the triangles with $\Delta t= 20
    \ms$. Note that the results depend on the structure of the input
    set. Whereas in the left case all sequences have some overlap, in the
    right case sequence $0$ and sequence $3$ are pretty much disjoint from
    the others.}
\end{figure}

A typical example for a network of $50$ neurons trained with $5$ sequences of
length $8$ is shown in Fig.\ \ref{figure8} and Fig.\ \ref{figure9}. There
are several important features to point out. First of all the recall never
comprises all $8$ neurons of the trained sequence but only up to $7$ active
neurons. This is however not a universal feature but rather a characteristic
of the global inhibition circuit shutting down the system after ca.\ $7$
spike occurrences, see Fig.\ \ref{figure8}. Furthermore, note that the
recall of the sequences speeds up toward the end of the sequence. This is
partly due to the fact that the integrate-and-fire neurons used here do not
have a finite rise time for their spikes which allows them to instantaneously
affect their postsynaptic neurons.

In a network with more realistic neurons one would expect that there is a
lower limit on the speed with which sequences can be recalled in the
system. Preliminary studies with realistic Hodgkin-Huxley type neurons show
this effect \cite{Nowotny2002}. It has clear advantages for maintaining the
correct order of recall in the system. The microscopic internal dynamics of
the neurons thus seems to be non-negligible for the macroscopic performance
of the system. This will be discussed in more detail in forthcoming work.

The quality of recall of sequences depends very much on the sequence and the
piece presented as a cue. This is however also no surprise because sequence
overlaps occur at certain neurons in the sequence and if these are used as a
cue the performance is less good as when other neurons are used. In Fig.\
\ref{figure9} one can see how some sequences are reproduced very well and
error free while others lead to activation of quite a few incorrect neurons.

\begin{figure}
\includegraphics[width=\columnwidth]{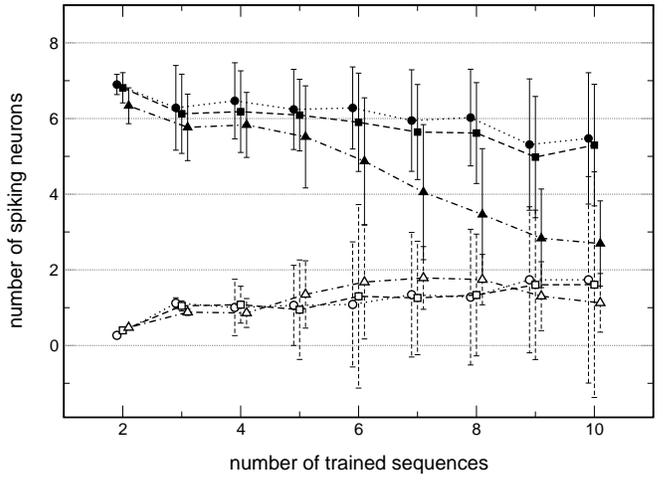}
  \caption{\label{figure10} Scaling of storage quality with the number of
    input sequences. A system with $50$ neurons was trained with a varying
    number of input sequences of length $8$. The figure shows the response
    after a total of $1600 \Delta t$ training for each input sequence. The
    filled symbols show the average number of responding neurons within a
    tested sequence and the open symbols the number of incorrectly responding
    neurons. The test cues were pieces of $2$ inputs from the trained
    sequences. The circles were obtained with sequences trained with inter
    spike intervals $\Delta t = 10 \ms$, the squares with $\Delta t= 15 \ms$,
    and the triangles with $\Delta t= 20 \ms$.  All data points are averages
    of trials with $5$ independently chosen sets of input sequences.}
\end{figure}

To test for the capacity of the system systematically we trained a network of
$50$ neurons with $2$ up to $10$ sequences of length $8$. For each number of
sequences $5$ independent sets of randomly chosen sequences were
tested. Fig.\ \ref{figure10} shows the average response of the trained
systems to pieces of $2$ inputs taken from the learned sequences. The
averages are over all possible subsequences and all $5$ input sequence sets
for each data point. This experiment was done with $3$ different input
speeds, i.e.\ the input was presented with fixed inter spike intervals of
length $\Delta t= 10 \ms$, $15 \ms$ and $20 \ms$. As one can see in
Fig.\ \ref{figure10} the performance dramatically decreases for the slowest
entrainment speed. This is due to the fact that the fixed width of the
learning window in (\ref{learneqn}) leads to weaker synapses for all the
synapses in this case as spikes are separated further in time, see last row
of Fig.\ \ref{figure7}. The minimum and maximum possible speed of the
entrainment are thus directly determined by the learning window. If one
chooses a larger learning window the slower sequences could be entrained as
well. However, this would also lead to decreased performance for faster
sequences.

\begin{figure}
\includegraphics[width=\columnwidth]{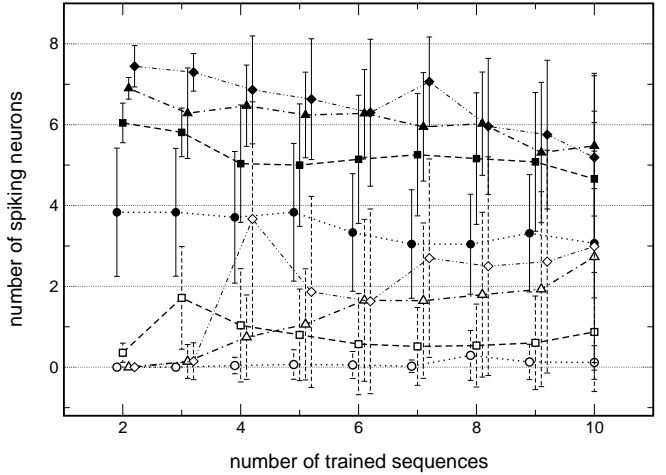}
  \caption{\label{figure11} Scaling of storage quality with the length of
    input sequences. A system with $50$ neurons was trained with sets of $5$
    input sequences of different lengths. The figure shows the response after
    a total of $16 \s$ training for each input sequence. The filled
    symbols show the average number of responding neurons within a tested
    sequence and the open symbols the number of incorrectly responding
    neurons. The test cues were pieces of $2$ inputs from the trained
    sequences. The circles were obtained with sequences of length $6$, the
    squares with length $7$, the triangles with length $8$, and the diamonds
    with length $9$. All data points are averages of trials with $5$
    independently chosen sets of input sequences.}
\end{figure}  

\begin{figure}
\includegraphics[width=\columnwidth]{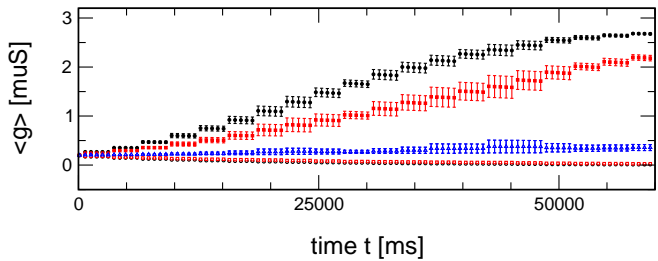}
\caption{\label{figure12} Development of synaptic strength during training
  of a sequence of length $6$ with speed $\Delta t= 10 \ms$. The network of
  $50$ neurons was trained with $5$ sequences of length $6$ in sequential
  order.  Each sequence was presented for $80 \Delta t$ at a time. The data
  shown are average synaptic strengths of synapses between the neurons of one
  of the trained sequences. The topmost points are the average strengths of
  all synapses between the neurons and their direct successors in the
  sequence, the middle are the corresponding strengths of synapses between
  neurons who are next nearest neighbors in the sequence under
  consideration, and the lower points correspond to strengths of synapses
  between neurons with distance $3$ in the sequence. Note how the synaptic
  strength for these synapses is suppressed because a spike being the third
  predecessor of a given spike is also the third successor of this spike due
  to cyclic training.  The lowest data points are the strengths between the
  neurons of the sequence as described above but {\em against} the order of
  activation in the trained sequence.}
\end{figure}

To test for the dependence of learning success on the length of presented
sequences we entrained a $50$ neuron system with sets of $5$ sequences of
length $6$ to $9$. Fig.\ \ref{figure11} shows the performance of the
system. On first sight it is surprising that the system performs less good
for shorter sequences. Naively one would expect a better performance because
overlaps are less likely. Indeed one really can see that the number of
erroneous spikes is smaller. On the other hand the number of correct spikes
is also considerably smaller. This is due to the fact that the spikes
preceding a given spike in a sequence are also succeeding it because of the
periodic presentation of the sequences (see e.g.\ Fig.\
\ref{figure3}). Synapses between the corresponding neurons are therefore
enhanced as well as depressed. For shorter sequences the last presentation of
the sequence is closer and the depression effect therefore stronger leading
to lesser overall synapse strength, cf Fig.\ \ref{figure12}. This creates
the fewer retrieved spikes for shorter sequences in Fig.\ \ref{figure11}. To
some extent this can be seen as an artifact because longer learning time or
slightly larger learning increments $A_+$ could diminish this effect. On the
other hand this might have negative effects on the performance of the system
in other parameter regions.

\section{Robustness}\label{noisesec}

Biological neural systems are subject to various external and internal noise
sources. Starting from internal thermal noise within the system this ranges
over noisy or unreliable input and influences from other parts of the organism
up to external electromagnetic fields. To test the effect of noise on the
learning success of our model systems we focused on two types of noise. We
implemented a Gaussian white noise in the membrane potential of the integrate-and-fire neurons
and we implemented unreliable input.

The internal white noise was added to the membrane potential of
each neuron independently. It is fully characterized by its mean, $0 \mV$ and
its variance for which several values between $0.2 \mV$ and $1.0 \mV$ were
tested. 

To simulate unreliable input we implemented Poisson input neurons. These
neurons produce rectangular spikes of width $t_{\text{spike}}= 3 \ms$ as
before but the time of spiking is stochastic. The spike times are determined
by the Poisson distribution
\begin{align}
  P(n_{\text{spike}} = k) = e^{- \lambda t} \frac{(\lambda t)^k}{k !}
  \end{align}
where $n_{\text{spike}}$ is the number of spikes occurring in an interval of
length $t$ and the parameter $\lambda$ is the mean firing rate. For small $t$
this can be approximated by $P(n_{\text{spike}} = 1) = \lambda t$,
$P(n_{\text{spike}} = 0) = 1- \lambda t$ and $P(n_{\text{spike}} = k) = 0$
for $k > 1$. This is the probability distribution we use to decide whether a
neuron fires within a time step of the Runge Kutta algorithm used. After
firing the neurons are refractory for $t_{\text{refract}}= 10 \ms$.  The
training protocol is that the mean firing rate of the first neurons is
switched from $0$ to some activity level $\lambda_{\text{on}}$ for $2 \Delta
t$, the next neuron is switched on after $\Delta t$ for also $2
\Delta t$ and so on.  Different reliability of the input can be adjusted by
the parameter $\lambda_{\text{on}}$.

\begin{figure}
  \includegraphics[width=\columnwidth]{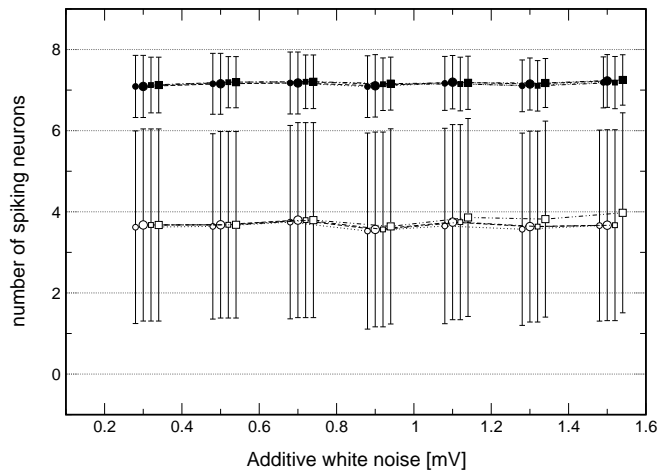}
  \caption{\label{figure13} Impact of Gaussian white noise in the membrane
    potential. The data points are the number of spiking neurons within
    tested sequences after $2400 \Delta t$ training at $\Delta t= 10 \ms$
    (full symbols) and the number of erroneously spiking neurons (open
    symbols).  The small symbols were obtained when the noise was only
    present during learning and the large ones when noise was always
    present. The circles correspond to a cue of two inputs in testing and the
    squares to a cue of three inputs.}
\end{figure}

\begin{figure}
  \includegraphics[width=\columnwidth]{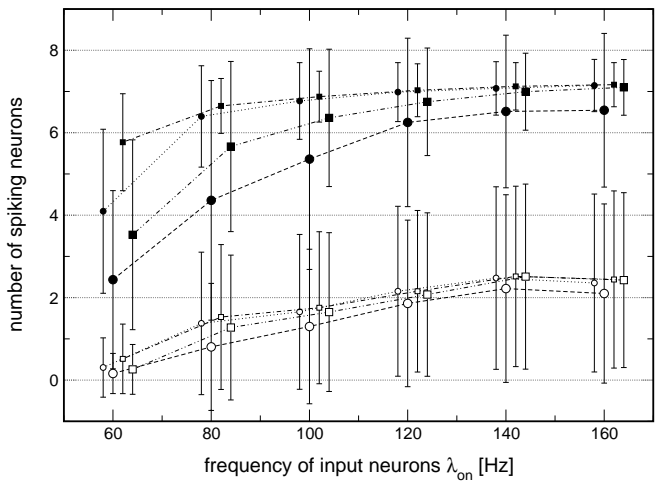}
  \caption{\label{figure14} Impact of noisy input on the learning
    performance. The input sequences were provided by stochastic Poisson
    neurons as described in the text. The data points are the number of
    spiking neurons within tested sequences after $2400 \Delta t$ training at
    $\Delta t= 10 \ms$ (full symbols) and the number of erroneously spiking
    neurons (open symbols).  The small symbols were obtained when the
    stochasticity of the input was only present during learning and the large
    ones when input was always stochastic. The circles correspond to a cue of
    two inputs in testing and the squares to a cue of three inputs.}
\end{figure}
Figs.\ \ref{figure13} and \ref{figure14} show the impact of the two types of
noise on the learning performance. Fig.\ \ref{figure13} shows the effect of
additive white noise at the membrane potential in the learning stage and in
both learning and recalling. As mentioned, the standard deviation of the
noise was chosen between $0.3 \mV$ and $1.5 \mV$. The system seems to
be more or less unaffected by noise of this magnitude. As to be expected the
learning is even less sensitive to noise than the recalling due to the fact
that the effect of the temporally uncorrelated noise on the synaptic strength
is averaged out over time.

Fig.\ \ref{figure14} shows the learning success if the input neurons fire
stochastically during learning only and during learning and recall as
described above. The parameter $\lambda_{\text{on}}$ was varied from $60$ to
$160 \, Hz$. The stochastic firing of the input neurons seem to only affect
the overall number of spikes, i.e.\ correct spikes as well as incorrect ones
but not their ratio. This indicates that mainly missing input spikes during
the training and especially during testing are responsible for the decreased
spikes in the response. It is to be expected that longer training can
diminish these effects even more. Like in the case of noise in the membrane
potential the learning stage is not affected as much by the noisy input as
the recall. Again the same argument applies. the effects of the stochasticity
of the input spikes is averaged out over time during the multiple repetitions
in the training phase.

\section{Discussion}
It has been demonstrated that STDP allows the transformation of temporal
information into spatial information providing an efficient mechanism for
storing temporal sequences which does not require a sophisticated network
topology. It is however not obvious how to {\em quantify} the storage
capacity of the system from the observed recall performance for different
numbers of stored sequences. Taking the heuristic rule to allow for
successful storage on average one incorrect spike in recall, the capacity of
a system of $50$ neurons is about $5-6$ sequences (see
Fig. \ref{figure10}. The capacity estimates for $n=50$ and $k=8$ are
$r(8,50,\frac{1}{2}) \approx 6.3$ and $\hat{r}(8,50,\frac{1}{2}) \approx
2.6$.  The storage capacity of the system therefore seems to be mainly
limited by the statistical properties of the input, i.e.\ the overlap
probabilities for randomly chosen input sequences. The biologically found
STDP learning rule obviously does not imply severe restrictions on
the ability to learn sequences but on the contrary seems to be very well
suited for this task. There are indications that the learning mechanism is
even more reliable with biologically more realistic conductance based model
neurons which have non-trivial intrinsic dynamics which to some extent
prevents the speedup in recall already discussed above.

The successful storage of arbitrary input sequences, however, crucially depends
on the existence of the corresponding synapses making the all-to-all
connections in the investigated system a necessary requirement. In realistic
systems such global all-to-all connections can not be found, but this might be
compensated through divergence and redundancy of the input. If the density of
connections and the number of neurons each input excites is high enough, pairs
of connected neurons being excited by successive inputs will appear on a
statistical basis. This mechanism will be discussed more thoroughly in
forthcoming work.

The realistic implementation of saturation of synaptic strength for additive
learning rules is another important topic. For the
system investigated here we implemented a combination of two mechanisms. On
the one hand the synaptic strength was directly bounded by use of the sigmoid
filtering function applied to the bare synaptic strength subject to the
additive learning rule, a technique commonly used by biologists. On the other
hand the steady decay of synaptic strength and the continuous stimulation of
the network by the inputs lead to a dynamical steady state thereby bounding
the synaptic strength dynamically.

Whereas the direct bound through a sigmoid filtering function might capture
some aspects of the behavior of real synapses, the decay of synaptic strength
necessary to achieve a realistic dynamical steady state is clearly too fast
to be realistic. The system forgets much too fast if it is not continuously
stimulated with appropriate input.

Alternative solutions to the saturation problem include competition based
mechanisms suggested by recent findings of interactions of various kinds
between neighboring synapses on a dendritic tree~\cite{Jensen2002} and
learning rules which depend on the synaptic strength itself like e.g.\
multiplicative learning rules.

The system is reasonably robust against noise. It is noteworthy that it is
not very sensitive to internal high-frequency noise. In the range of noise
applied in our trials the recall barely depended on the level of noise (see
Fig. \ref{figure13}. Whether this is an effect of the integrate-and-fire
neuron model used here is beyond the scope of this work. The
tolerance to biologically more relevant noise in the spike timing of the
input is also rather impressive taking into account that $\lambda_{\text{on}}
= 60 \, Hz$ corresponds to a total firing probability of only $36 \%$ for
each of the input neurons within their activity window of $20
\ms$. Nevertheless the system still was able to store at least parts of the
presented sequences at this high noise level.

\section*{Acknowledgments}
We thank Walter Senn for numerous helpful remarks and suggestions.
This work was partially supported by the U.S. Department of Energy, Office of
Basic Energy Sciences, Division of Engineering and Geosciences, under Grants
No. DE-FG03-90ER14138 and No. DE-FG03-96ER14592, by grants from the National
Science Foundation, NSF PHY0097134 and NSF EIA0130708, by a grant from the
Army Research Office, DAAD19-01-1-0026, by a grant from the Office of Naval
Research, N00014-00-1-0181, and by a grant from the National Institutes of
Health, NIH R01 NS40110-01A2.

\begin{appendix}
  \newcommand{\ts}{\textstyle}
  \section{Taylor expansion of $\mathbf p_i^j$} \label{appe}
 We first need to proof the identity
\begin{widetext}
  \begin{align}
    \frac{d^n}{dx^n} {\ts \binom{r}{s} x^s (1-x)^{r-s}} = \sum_{k=
    \max\{n+s-r,0\}}^{\min\{s,n\}} {\ts \binom{r}{s} 
    \frac{s!}{(s-k)!} \frac{(r-s)!}{(r-s-(n-k))!} \binom{n}{k} (-1)^{n-k}
    x^{s-k}(1-x)^{r-s-(n-k)}} . \label{conjecteqn}
  \end{align}
\end{widetext}
  The proof is by induction. Let $n=0$. Then the equation reduces to
  \begin{align}
    \ts \binom{r}{s} x^s (1-x)^{r-s} = \binom{r}{s} \frac{s!}{s!}
    \frac{(r-s)!}{(r-s)!} \binom{0}{0} (-1)^0 x^s (1-x)^{r-s}
  \end{align}
  which is clearly true. Assuming the validity of (\ref{conjecteqn}) for $n$
  we can calculate
\begin{widetext}
  \begin{align}
    & \frac{d^{n+1}}{dx^{n+1}} {\ts \binom{r}{s} x^s (1-x)^{r-s}} \\
    &= \frac{d}{dx} \bigg(\sum_{k=\max\{n+s-r,0\}}^{\min\{s,n\}} {\ts
    \binom{r}{s} 
    \frac{s!}{(s-k)!} \frac{(r-s)!}{(r-s-(n-k))!} \binom{n}{k} (-1)^{n-k}
    x^{s-k} (1-x)^{r-s-(n-k)}} \bigg) \\
    &= \sum_{k=\max\{n+s-r,0\}}^{\min\{s,n\}} {\ts \binom{r}{s}
    \frac{s!}{(s-k-1)!} \frac{(r-s)!}{(r-s-(n-k))!} \binom{n}{k} (-1)^{n-k}
    x^{s-k-1} (1-x)^{r-s-(n-k)} } \\
    &\hphantom{=} + \sum_{k=\max\{n+s-r,0\}}^{\min\{s,n\}} {\ts \binom{r}{s}
    \frac{s!}{(s-k)!} \frac{(r-s)!}{(r-s-(n+1-k))!} \binom{n}{k} (-1)^{n+1-k}
    x^{s-k}(1-x)^{r-s-(n+1-k)}} .
  \end{align}
\end{widetext}
  Shifting the index in the first sum by one, using the well known
  identity $\binom{n}{k}+\binom{n}{k-1}= \binom{n+1}{k}$ and obvious
  identities like $1= \binom{n+1}{0}$ one obtains equation (\ref{conjecteqn})
  for $n+1$ which completes the proof.

  The Taylor expansion for $p_j^i$ is then straightforward:
\begin{widetext}
  \begin{align}
    p_j^i &= 1 - \sum_{s=0}^{i-1} {\ts \binom{r}{s} p_j^s (1-p_j)^{r-s}} \\
    &= - \sum_{n=1}^{\infty} \sum_{s=0}^{i-1} \bigg(\sum_{k=
    \max\{n+s-r,0\}}^{\min\{s,n\}} {\ts \binom{r}{s} 
    \frac{s!}{(s-k)!} \frac{(r-s)!}{(r-s-(n-k))!} \binom{n}{k} (-1)^{n-k}
    p_j^{s-k}(1-p_j)^{r-s-(n-k)}} \bigg) \bigg|_{p_j=0} \frac{(p_j)^n}{n!} .
  \end{align}
\end{widetext}
  For all $k < s$ the $n$-th derivative contains a non-zero power of $p_j$
  and is thus $=0$ at $p_j= 0$. Furthermore, if $s>n$ then all $k$ are less
  then $s$ and therefore the whole sum over $k$ is empty. We end up with
  \begin{align}
    p_j^i &= - \sum_{n=1}^{\infty} \sum_{s=0}^{\min\{i-1,n\}} {\ts \binom{r}{s}
    \frac{s!(r-s)!}{(r-n)!} \binom{n}{s} (-1)^{n-s}} \frac{(p_j)^n}{n!} \\
    &= -\sum_{n=1}^{\infty} \sum_{s= 0}^{\min\{i-1,n\}} {\ts \binom{r}{n}
    \binom{n}{s} (-1)^{n-s} (p_j)^n } .
  \end{align}
  For any $n \leq i-1$ the inner sum is
  \begin{align}
     {\ts \binom{r}{n} (p_j)^n} \sum_{s=0}^{n} {\ts \binom{n}{s}
     (-1)^{n-s} 1^s=  {\ts \binom{r}{n} (p_j)^n} (1-1)^ n
     = 0} .
    \end{align}
  Therefore, the leading term of the Taylor expansion of $p_j^i$ is
  \begin{align}
    p_j^i = \binom{r}{i} (p_j)^i + {\cal O}((p_j)^{i+1}).
  \end{align}

  \end{appendix}

\end{document}